\documentclass[a4paper,11pt]{article}
\usepackage{jheppub} 
\usepackage{lineno}

\title{\boldmath Gravitational waves and dark matter with witten effect}
\author{Ruiyu Zhou$^{a,b}$}
\emailAdd{zhoury@cqupt.edu.cn}
\author{Ligong Bian$^{b}$\footnote{Corresponding Author.}}
\emailAdd{lgbycl@cqu.edu.cn}

\affiliation{
$^a$~School of Science, Chongqing University of Posts and Telecommunications, Chongqing 400065, P. R. China\\
$^b$~Department of Physics and Chongqing Key Laboratory for Strongly Coupled Physics, Chongqing University, Chongqing 401331, P. R. China
}

\abstract{
We investigate the breaking of dark $SU(2)_d$ symmetry at different temperature scales, occurring after Peccei-Quinn symmetry breaking or following QCD symmetry breaking. 
We focus on assessing the potential of the hidden monopoles generated during this process to serve as dark matter candidate. 
Additionally, we examine the impact of axion-monopole interactions on the axion mass.
When the phase transition occurs at extremely high temperature ($\sim 10^8 \mathrm{GeV}$), the contribution of monopoles to the axion mass through witten effect becomes non-negligible, playing a crucial role in accurately determining the axion relic density. Moreover, the stochastic gravitational wave background generated by dark phase transition and axionic domain wall collapse may offer a potential explanation for the low-frequency gravitational wave signals observed in PTA experiments.}

\begin{document}
\maketitle

\section{Introduction} 
The successful discovery of gravitational waves has inaugurated a new era in gravitational wave astronomy and provided a novel approach to explore new physics beyond the Standard Model (SM)~\cite{Caldwell:2022qsj}. Many significant phenomena in the early universe, such as the electroweak phase transition and sphaleron, remain challenging to probe directly due to their occurrence at extremely high energy scales. However, detecting the gravitational wave signals associated with these events offers a promising avenue to deepen people's understanding of these phenomena. Moreover, considerable research efforts~\cite{Dev:2016feu,Machado:2019xuc,DelleRose:2019pgi,VonHarling:2019rgb} have focused on utilizing gravitational wave observatories, such as LIGO~\cite{LIGOScientific:2019vic} and ET~\cite{Punturo:2010zz}, to probe gravitational wave signals originating from the Peccei-Quinn(PQ) phase transition, offering a novel pathway for axion detection. 
Recently, pulsar timing arrays (PTAs)~\cite{EPTA:2023fyk,Reardon:2023gzh,NANOGrav:2023gor,Xu:2023wog} reported a potential detection of a low-frequency stochastic gravitational wave background, sparking considerable interest in exploring its origin, with nanohertz phase-transition gravitational waves~\cite{Bian:2020urb,Nakai:2020oit,Addazi:2020zcj,Ratzinger:2020koh,Borah:2021ftr,Xue:2021gyq,Chen:2023bms,He:2023ado,Wu:2023hsa,Athron:2023mer,Addazi:2023jvg} becoming a focal point of study.

The axion~\cite{Peccei:1977hh,Peccei:1977ur} was proposed as one of the most promising solutions to the long-standing strong CP problem in particle physics. As a pseudo-Nambu-Goldstone boson in the PQ symmetry breaking, the axion dynamically cancels the CP phase, providing an elegant solution to the strong CP problem. 
Moreover, the mass imparted to the axion by the QCD instanton effects~\cite{Weinberg:1977ma,Wilczek:1977pj,tHooft:1976snw} makes it one of the compelling candidates for cold dark matter (DM) through the misalignment mechanism\cite{Preskill:1982cy}. Within this framework, the cosmological abundance of the axion is governed by its mass (or its decay constant $f_a$, which is closely related to the axion mass and PQ symmetry breaking scale) and the initial misalignment of the axion field, typically expressed as $\theta_{mis}$. 
Non-perturbative QCD effects are not the only source of axion mass. Refs.~\cite{Fischler:1983sc,Julia:1975ff,Jeong:2023gjc,Banerjee:2024ykz} have demonstrated that the Witten effect can contribute mass to the axion and significantly influence its dynamics. 

The detection of magnetic monopoles has long been a central focus in physics and astronomy.
Recently, refs.~\cite{Zhang:2024mze,Bai:2021ewf,IceCube:2015agw,PierreAuger:2016imq} (see ref.~\cite{Mavromatos:2020gwk} for a recent review) have proposed novel detection methods or experimental constraints for magnetic monopoles.
In addition, the potential role of monopoles as topological DM has been extensively explored in refs.~\cite{Nakagawa:2021nme,Graesser:2021vkr,Fan:2021ntg,Graesser:2020hiv,Daido:2019tbm,Sato:2018nqy,Kawasaki:2015lpf,Hiramatsu:2021kvu,GomezSanchez:2011orv,Nomura:2015xil,Murayama:2009nj,Evslin:2012fe,Terning:2019bhg}. 
Furthermore, refs.~\cite{Baek:2013dwa,Khoze:2014woa,Bai:2020ttp} suggest that if the ’t Hooft-Polyakov monopole~\cite{tHooft:1974kcl,Polyakov:1974ek} originates from the dark sector, which could significantly contribute to the observed relic density. Additionally, the decay of topological defects, such as magnetic monopoles and domain walls, can generate axions, affecting their cosmological abundance. 
Moreover, ref.~\cite{Kawasaki:2015lpf} demonstrates that when considering the coupling between monopoles and QCD axions, the Witten effect on the axion abundance becomes non-negligible, particularly before the QCD phase transition. And, ref.~\cite{Banerjee:2024ykz} illustrates the key role of the interaction between axions and monopoles in alleviating the three-way tension in the pre-inflationary QCD axion DM scenario.  
This paper investigates the scenario in which the PQ symmetry is broken after cosmic inflation, known as the post-inflationary axion scenario.
Under these conditions, spontaneous breaking of the discrete subgroup $Z(N)$ of the $U(1)_{PQ}$ symmetry leads to the formation of topological defects known as domain walls \cite{Vilenkin:1982ks}. However, the presence of these domain walls can significantly affect the evolution of the universe, a phenomenon referred to as the domain wall problem~\cite{Zeldovich:1974uw,Sikivie:1982qv}. To address this problem, it is generally necessary for the domain walls to collapse before they overclose the universe~\cite{Vilenkin:1981zs,Gelmini:1988sf,Larsson:1996sp}. This collapse process also involves gravitational wave radiation, and detecting these stochastic gravitational wave background signals can provide significant constraints on axion models.

Given the rich phenomenology associated with the $SU(2)_d$ dark sector, a substantial body of literature has already explored cosmological phase transition~\cite{Baldes:2018emh,Hall:2019ank,Ghosh:2020ipy,Borah:2021ftr}, DM~\cite{Fornal:2017owa,Borah:2021ftr,Ghosh:2020ipy,Hambye:2008bq,Boehm:2014bia,Gross:2015cwa,Ko:2020qlt,Chen:2015nea,Chen:2015dea,Chiang:2013kqa}, and hidden monopole~\cite{Daido:2019tbm,Baek:2013dwa,Khoze:2014woa}.
In this paper, we focus on the hidden monopoles generated by the collisions of vacuum bubbles during a first-order phase transition and the impact of the interaction between hidden monopoles and axions on the axion mass and relic density. We also estimate the stochastic gravitational wave background produced by dark phase transition and domain wall collapse.

This paper is organized as follows. In Sec.~\ref{model}, we overview the dark $SU(2)_d$ model and the finite temperature effective potential. The Witten effect and the relic density of both monopole and axion will be explored in Sec.~\ref{witteneffect}. Sec.~\ref{gw} discusses the gravitational wave from the dark phase transition and the domain wall collapse. Our summary is drawn in Sec.~\ref{summary}.

\section{The dark $SU(2)_d$ phase transition}\label{model}
The Lagrangian of the dark gauge $SU(2)_d$ symmetry triplet scalar with the PQ symmetry complex scalar:
\begin{align}\label{larg}
\mathcal{L} &= \frac{1}{2} \partial_\mu \varphi^* \partial^\mu \varphi - \frac{1}{4} \lambda (|\varphi|^2 - v_\varphi^2)^2 - \frac{\lambda}{6}T^2|\varphi|^2 \nonumber \\
&-\frac{1}{4} F^a_{\mu\nu} F^{\mu\nu a} + \frac{1}{2}  D_\mu \phi_a D^\mu \phi_a  - \frac{\lambda_\phi}{4} (\phi_a\phi_a)^2 - \lambda_{\varphi \phi} |\varphi|^2 \phi_a\phi_a\;,
\end{align}
where $\varphi$ is a complex singlet scalar of the global $U(1)_{PQ}$ symmetry, $\phi_a$ is a dark $SU(2)_d$ triplet scalar, $\lambda$ is the PQ coupling constant, and $\lambda_{\varphi\phi}$ is the coupling between the $\varphi$ and $\phi$ . Therein, the field strength of the $SU(2)_d$ field $A_\mu$ is $F_{\mu\nu}^a = \partial_\mu A_\nu^a - \partial_\nu A_\mu^a + g_d \epsilon^{abc} A_\mu^b A_\nu^c$, the kinetic term is $D_\mu \phi_a = \partial_\mu \phi_a + g_d \epsilon^{abc} A_\mu^b \phi_c$, with $g_d$ being the gauge coupling and $\phi_a (a= 1,2,3)$ being the adjoint scalars.
As the universe expands, $\varphi$ acquires a non-zero vacuum expectation value (VEV), leading to spontaneous breaking of the PQ symmetry, which is $\langle \varphi \rangle = v_\varphi$. Subsequently, the dark $SU(2)_d$ symmetry spontaneously breaks, with the triplet acquiring a non-zero VEV, $\langle\phi_3\rangle = v_\phi $. After symmetry breaking of $SU(2)_d \to U(1)_d$, there are two massive charged gauge bosons $W^\prime_{\pm}$ with $m_{W^\prime} = g_d v_\phi$, one massless gauge boson $\gamma^\prime$, and one massive scalar $\phi$ with $m_{\phi} = -2 \lambda_{\varphi \phi} v_\varphi^2 + 3 \lambda_\phi v_\phi^2$. In addition, the hidden ’t Hooft-Polyakov monopole~\cite{tHooft:1974kcl} may be formed.

This study explores the breaking of dark $SU(2)_d$ symmetry at different phase transition temperature scales (the sub-EeV and sub-GeV levels), which are determined by the VEV of PQ symmetry $v_{\varphi}$ and the coupling $\lambda_{\varphi \phi}$. For simplicity in the discussion, we fixed the parameter $v_\varphi$ to $10^{10}$ GeV in this study.
The nature of the dark sector requires its coupling with the visible sector to be extremely weak. In this paper, we set the coupling constant between $\phi_3$ and the axion $\lambda_{\varphi \phi} \ll 1$.
To study the phase transition of dark $SU(2)_d$ symmetry, we use the standard approach by employing the thermal one-loop effective potential~\cite{Quiros:1999jp}
\begin{align}
    V_{\text{eff}}(\phi_3,T)=V_\text{{tree}}(\phi_3)+V_\text{{CW}}(\phi_3)+V_\text{{c.t}}(\phi_3)+V_1^\text{T}(\phi_3,T)+V_1^\text{{daisy}}(\phi_3,T).
\end{align}
The $V_{\text{eff}}(\phi_3,T)$ and $V_\text{CW}(\phi_3)$ are the tree-level potential and the 1-loop Coleman-Weinberg potential, with $V_\text{c.t}(\phi_3)$ to keep the zero temperature vacuum un-shifting. The finite temperature correction is described by the term $V^T_1(\phi_3,T)$ and the daisy correction term $V^{daisy}_1 (\phi_3,T)$. The specific form of each thermal correction can be found in the Appendix~\ref{thermal-correction}.

The characteristic temperatures of the phase transition distinguish the different stages of this process, which is crucial to understand the dynamics of the transition. The critical temperature $T_c$ is the temperature at which the system exhibits two degenerate vacuum states in co-existence. Bubble nucleation occurs at $T_n$, where, on average, one bubble is nucleated within one unit horizon volume. At percolation temperature $T_p$, the probability of finding a point remaining in the false vacuum is $70\%$. At this time, it can be roughly considered as the completion of the phase transition.

The three characteristic temperatures can be determined accordingly using the definitions of these characteristics outlined above.
The bubble nucleation rate $\Gamma$ is defined by~\cite{Wang:2020jrd} $\Gamma \sim T^4 e^{-S_3/T} \left( S_3/{2 \pi T}\right)^4 $. 
And the definition of the action of the bubble solution $S_3$ is given by
\begin{eqnarray}
S_3 = 4 \pi \int_0^{\infty} dr r^2 \left[ \frac{1}{2} \left(\frac{d \phi}{d r}\right)^2 + V_{\text{eff}}(\phi , T) \right],
\end{eqnarray}
where $\phi(r) = \phi_3 $ is the ``bounce solution'' acquired from the equations of motion
\begin{eqnarray}
\frac{d^2 \phi}{d r^2} + \frac{2}{r} \frac{d \phi}{d r} = \frac{\partial V_{\text{eff}}}{\partial \phi},
\end{eqnarray}
with the boundary conditions
\begin{eqnarray}
\left. \frac{d \phi}{d r} \right|_{r=0}= \phi_{\text{phase 1}}, ~~\left. \frac{d \phi}{d r} \right|_{r=\infty}= \phi_{\text{phase 2}}, 
\end{eqnarray}
between the two phases $\phi_{\text{phase 1}}$ (true vacuum) and $\phi_{\text{phase 2}}$ (false vacuum) during the transition.
In this work, we use the FindBounce~\cite{Guada:2020xnz} package to calculate the ``bounce solution''.
The probability of a point remaining in the false vacuum is defined as~\cite{Leitao:2012tx, Guth:1979bh, Guth:1981uk}:
\begin{eqnarray}
P(T) = \exp\left[ \frac{4\pi v_w^3}{3} \int_T^{T_c} \frac{dT' \, \Gamma(T')}{H(T') T'^4} \left( \int_T^{T'} \frac{d\tilde{T}}{H(\tilde{T})} \right)^3\right]\;,
\end{eqnarray}
where $v_w$ is the velocity of the bubble wall.
\begin{figure}
    \centering
    \includegraphics[width=0.45\linewidth]{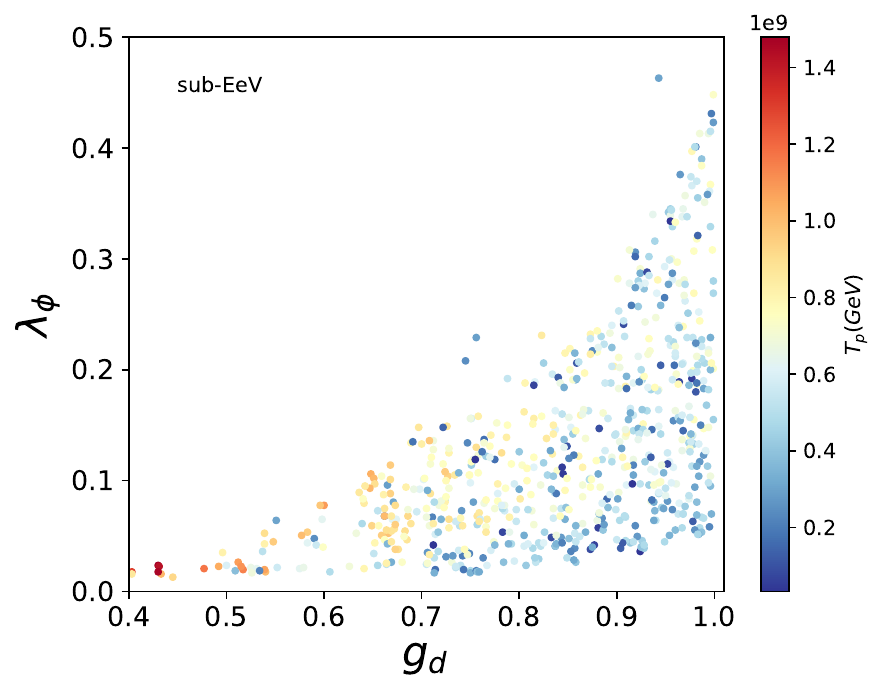}
    \includegraphics[width=0.46\linewidth]{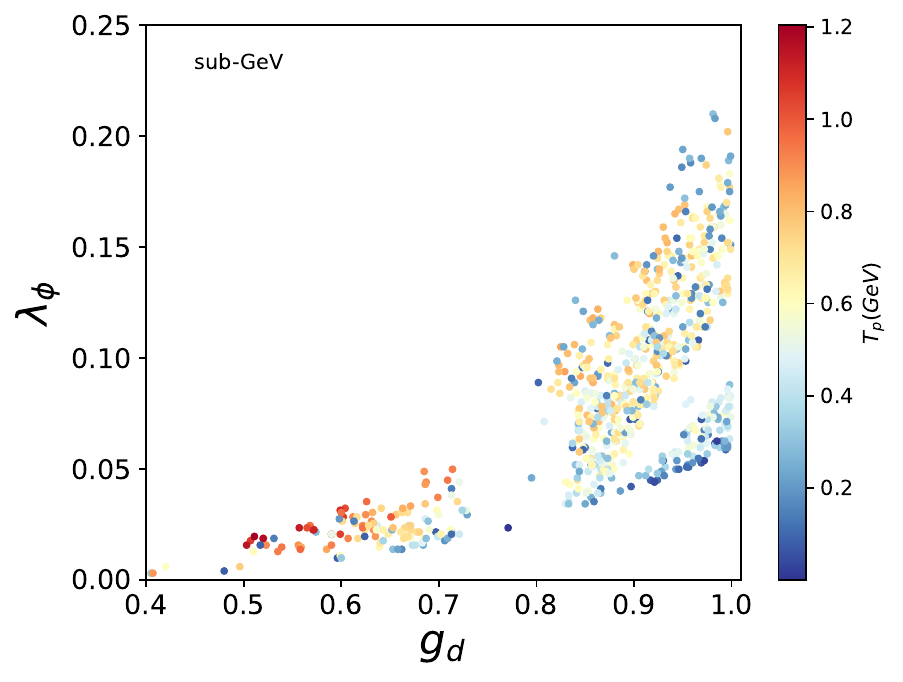}
    \caption{The left and right panels illustrate the distribution of viable parameter points in the $\lambda_\phi - g_d$ plane corresponding to phase transitions occurring at the sub-EeV and sub-GeV energy scales, respectively. The colormap on the right side of the figure represents the percolation temperature $T_p$ of the phase transition.}
    \label{fig:couplingFOPT}
\end{figure}

By simultaneously varying $v_\varphi$ and $\lambda_{\varphi \phi}$ while keeping the product $\lambda_{\varphi \phi} (T^2 + 3v_\varphi^2)$ constant, the feasible parameters of the phase transition can be matched to scenarios with different axion decay constant $f_a$. 
Through numerical calculations, we present the viable parameter points for the first-order phase transition in Fig.~\ref{fig:couplingFOPT}, where the points in the left (right) figure represent the phase transition occurring around the temperature $T_p \sim \mathcal{O}(10^8)$ GeV ($ T_p \sim \mathcal{O}(10^{-1})$ GeV).
As shown in Fig.~\ref{fig:couplingFOPT}, the viable parameter space for the sub-EeV first-order phase transition is primarily concentrated in regions with large gauge couplings $g_d > 0.4$ and small scalar self-couplings $\lambda_\phi < 0.45$. In addition, it is observed that the percolation temperature $T_p$ tends to decrease as $g_d$ increases. In contrast, for the sub-GeV phase transition, the viable parameter space favors smaller scalar self-couplings $\lambda_\phi < 0.2$.

\section{The Witten effect and the DM relic density}\label{witteneffect}

During the phase transition, monopoles can form at a rate of approximately $p \sim \mathcal{O}(10^{-1})$ as vacuum bubbles collide with each other. 
Thus, a quantitative relationship can be established to express the number density of monopole $n_{M}$ in terms of the number density of colliding bubbles $n_{b}$, which also demonstrates the crucial role of bubble collisions in the monopole production process~\cite{Einhorn:1980ik},
\begin{align}
n_{M} = p n_{b} = p \frac{\beta^3}{8\pi v_w^3} .
\end{align}

Hidden monopoles are formed when the $SU(2)_d$ symmetry spontaneously breaks down to $U(1)_d$. Then, the configurations of the scalar and gauge fields are presented as follows~\cite{Hiramatsu:2021kvu}:
\begin{align}
\phi_a=v_\phi H(r)\frac{x_a}{r}\;,   \  \
A_{i}^{a}=\frac{1}{g_d}\frac{\epsilon^{aij} x
^j}{r^2}F(r)\;,   \  \  (i,j=1,2,3),
\end{align}
where $\epsilon^{aij}$ is the totally antisymmetric tensor of rank $3$ with a convention $\epsilon^{123}=1$, and $r=\sqrt{x^2+y^2+z^2}$.
In terms of the above configurations and dimensionless variable $\xi = v_\phi r$, the Lagrangian given in Eq.~(\ref{larg}) reduces to
\begin{align}
L=\frac{4\pi v_\phi}{g_d^2} \int_{0}^{\infty} d\xi &\bigg[\left(\frac{dF}{d\xi}\right)^2+\frac{2F^2(1-F)^2}{ \xi^2} +\frac{F^4}{2 \xi^2}+\frac{g_d^2 \xi^2}{2} \left(\frac{dH}{d\xi} \right)^2+g_d^2H^2(1-F)^2 \nonumber \\ &+\frac{g_d^2 V_{eff}(H)}{v_\phi^4}\xi^2 \bigg].
\end{align}
The boundary conditions for the aforementioned equations of motion can be written as follows: 
\begin{align}\label{eom}
&\frac{d^2F}{d\xi^2}=\frac{F}{\xi^2}(1-F)(2-F)+g_d^2H^2(F-1)\;,\\
&\frac{d^2H}{d\xi^2}+\frac{2}{\xi}\frac{dH}{dx}=\frac{2H}{\xi^2}(1-F)^2+\frac{1}{v_\phi^4}\frac{dV_{eff}(H)}{dH}\;.
\end{align}
Here, the functions $H(\xi)$ and $F(\xi)$ satisfy boundary conditions:
\begin{align}
\lim_{\xi\to 0}H(\xi) \rightarrow 0\;,  \lim_{\xi\to 0 }F(\xi) \rightarrow 0\;, \lim_{\xi \to \infty} H(\xi) \rightarrow 1\;, \lim_{\xi\to \infty }F(\xi) \rightarrow 1\;.  
\end{align}

\begin{figure}[htbp]
    \centering
    \includegraphics[width=0.6\linewidth]{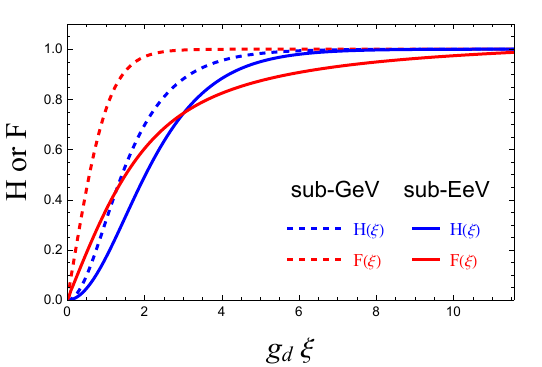}
    \caption{The blue and red curves represent the profiles of $H(\xi)$ and $F(\xi)$, respectively, where the horizontal axis corresponds to $g_d \xi = r m_{w^\prime} $. The solid, and dashed lines correspond to BM1 and BM4 in the Table.~\ref{tab:BP} respectively.}
    \label{fig:handf}
\end{figure}
In Fig.~\ref{fig:handf}, we use BM1 from the sub-GeV scale and BM4 from the sub-EeV scale to illustrate the profiles of $H(\xi)$ and $F(\xi)$ at the percolation temperature $T_p$. The horizontal axis in this figure is chosen as $g_d \times  \xi = r \times m_{w^\prime}$ to represent the core size of a monopole.
With $H(\xi)$ and $F(\xi)$ being calculated numerically, the monopole mass can be calculated using the following equation, 
\begin{align}\label{mm}
m_M=&\frac{4\pi v_\phi}{g_d^2} \int_{0}^{\infty} d\xi \bigg[\left(\frac{dF}{d\xi}\right)^2+\frac{2F^2(1-F)^2}{ \xi^2} +\frac{F^4}{2 \xi^2}+\frac{g_d^2 \xi^2}{2} \left(\frac{dH}{d\xi} \right)^2 \nonumber \\ 
+&g_d^2 H^2(1-F)^2
+\frac{g_d^2 V_T(H)}{v_\phi^4}\xi^2-\frac{g_d^2 V_T(H=1)}{v_\phi^4}\xi^2 \bigg].
\end{align}

The attractive interaction between monopoles and antimonopoles leads to annihilation, a process known as “diffusive capture” \cite{Preskill:1979zi,Goldman:1980sn,Vilenkin:2000jqa,Zeldovich:1978wj}, which further impacts the monopole number density $n_M$. This process can take place only if the mean free path of the monopole, $\bar\lambda_{mfp} \sim \sqrt{m_M/T}/(BT)$, is shorter than its capture radius, $r_{cap} \sim h^2/T$ \cite{tHooft:1974kcl,Polyakov:1974ek}, where $B$ represents the drag force exerted on the monopole by the thermal plasma \cite{Preskill:1979zi,Goldman:1980sn,Vilenkin:2000jqa}, and $h=1/g_d$ denotes the hidden magnetic charge of the monopole. 
The situation we study is the same as that in \cite{Banerjee:2024ykz}, where $\bar\lambda_{mfp}>r_{cap}$, meaning that the annihilation of monopoles with anti-monopoles does not affect their number density $n_M$ generated during the phase transition.
Once the monopole number density $n_M$ and mass $m_M$ are determined, its relic density can be calculated using the definition of the density parameter, as expressed below: 
\begin{align}
    \Omega_M h^2 = \frac{\rho_{M,0} }{\rho_{\mathrm{crit},0}} h^2 = pm_M\frac{\beta^3g_s(T_0)T_0^3h^2}{24\pi v_w^3 g_s(T_*)T_*^3M_{pl}^2H_0^2} \;,
\end{align}
where, $s_{*}$($s_{0}$) represents the entropy density of the universe at the end of phase transition(today), $g_s$ is the effective number of degrees of freedom in entropy, and $H_0$ denotes the present-day Hubble constant.

\begin{figure}
    \centering
    \includegraphics[width=0.45\linewidth]{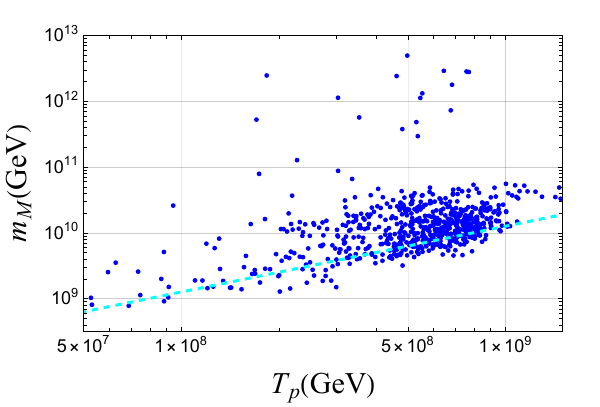}    \includegraphics[width=0.46\linewidth]{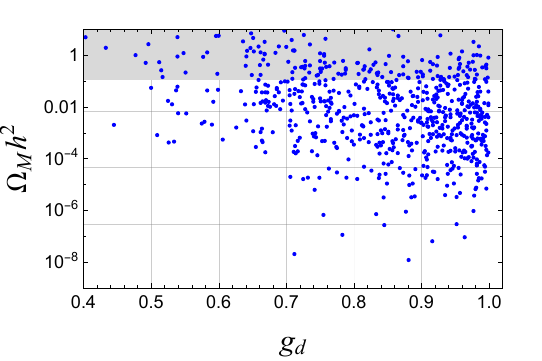}
    \caption{The left panel shows the relationship between the monopole mass $m_M$ and the percolation temperature $T_p$. The cyan dashed line represents the value $4\pi T_p$, providing a reference scale for comparison. The right panel illustrates the evolution of the monopole relic density, $\Omega_M h^2$, as a function of $g_d$. The gray-shaded region represents the constrain of observed DM abundance~\cite{Planck:2018vyg}}
    \label{fig:mMOmega}
\end{figure}
As discussed above, the mass and relic density of the monopole are closely related to the energy scale of the phase transition. In the sub-GeV case, the monopole mass and relic density are extremely minimal, rendering them unobservable. Therefore, in the subsequent discussion, we focus solely on the scenario in which the phase transition occurs at the sub-EeV scale.
The left panel of Fig.~\ref{fig:mMOmega} presents the relation between percolation temperature $T_p$ and monopole mass $m_{M}$ calculated using the formulas mentioned above.
From this figure, it can be observed that the monopole mass is approximately proportional to the percolation temperature $T_p$, following the relation $m_M \sim 4\pi v_\phi / g_d^2 \sim 4 \pi T_p$ since most points are concentrated around $v_\phi(T_p) \sim T_p$ with $g_d^2 \sim 1$. Furthermore, the right panel of Fig.~\ref{fig:mMOmega} shows that most viable points from the sub-EeV phase transition contribute non-negligibly to the DM relic density, whereas some of the heavier monopoles are excluded by the current observed relic density~\cite{Planck:2018vyg}.

In our investigation of the influence of the Witten effect on axion mass, we utilize the coupling framework consistent with that presented in~\cite{Fischler:1983sc}, which describes the interaction between the axion and the dark $U(1)_d$ gauge field.
\begin{align}
\mathcal{L}_\theta = -\frac{e^{\prime2}}{32 \pi^2}\frac{a}{f_a}F^{\prime\mu\nu} \tilde{F}^\prime_{\mu \nu}\;,
\end{align}
where $e^\prime \equiv 4 \pi /  g_d$ is the dark electric gauge coupling. This coupling is inherited following the interaction between the axion and the dark $SU(2)_d$ gauge field.
Moreover, hidden magnetic monopoles acquire hidden electric charge from this Lagrangian, transforming into particles that simultaneously possess both electric and magnetic charges, known as dyons, and this effect is called the Witten effect~\cite{Witten:1979ey}. 
Furthermore, considering axions in an electromagnetic field that contains both monopoles and antimonopoles, the axion acquire an effective mass $m_{a,M}$, which is determined by the number density of monopoles~\cite{Fischler:1983sc,Kawasaki:2017xwt}
\begin{align}
    m_{a,M}^2 \simeq 2\beta \frac{n_M(T)}{f_a}\;,
    \label{eq:axionmass}
\end{align}
where $\beta \equiv e^{\prime 2}/{128\pi^3 r_c f_a}$, $n_M$ is the number density of the monopole and antimonopole, \textit{$n_M = n_M^+ + n_M^-$} and  
$r_c$, the core size of the hidden monopole, can be estimated as the inverse of the mass of the charged gauge bosons generated after the spontaneous symmetry breaking of the $SU(2)_d$ gauge symmetry.
\begin{figure}
    \centering
    \includegraphics[width=0.6\linewidth]{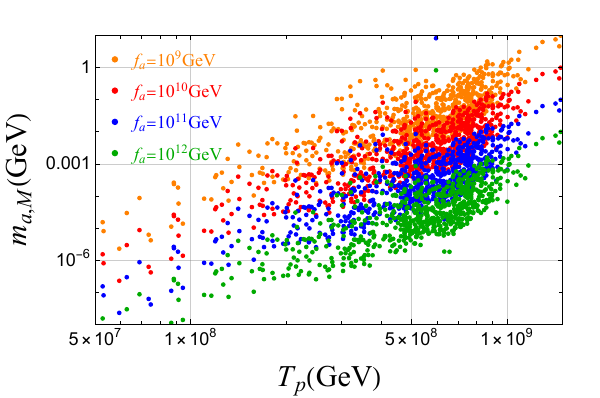}
    \caption{This figure illustrates the relationship between the percolation temperature $T_p$ and the axion mass $m_{a,M}$ obtained via the Witten effect. The orange, red, blue, and green points represent cases where the axion decay constant $f_a=10^{9}$ GeV, $f_a=10^{10}$ GeV, $f_a=10^{11}$ GeV, and $f_a=10^{12}$ GeV, respectively.}
    \label{fig:axionmass}
\end{figure}
In Fig.~\ref{fig:axionmass}, we present the relationship between the axion mass $m_{a,M}$ induced by the Witten effect and the percolation temperature $T_p$. As shown in Eq.(\ref{eq:axionmass}), the axion mass $m_{a,M}$ is inversely proportional to the axion decay constant $f_a$, with most $m_{a,M}$ clustering around $10^{-3}$ GeV. In the sub-EeV scenario, the contribution of the Witten effect to the axion mass significantly exceeds that of the QCD effect $m_{a,QCD} \sim 5.70 \rm{\mu eV}\times10^{16}\rm{GeV}/f_a$~\cite{GrillidiCortona:2015jxo,Hook:2018dlk}.

In the post-inflation scenario, the breaking of global PQ symmetry may lead to the formation of topological defect, such as global strings, which release energy through radiating free axions~\cite{Yamaguchi:1998gx,Hiramatsu:2010yu}. After acquiring mass through the Witten effect, the axion can be considered as cold DM.
Furthermore, we refer to the study in ref.~\cite{Jia:2024ejr} to explore the contribution of KSVZ axion~\cite{Kim:1979if,Shifman:1979if} (for $N_{DW} =1 $) produced by cosmic string radiation to the axion relic density, which can be written as:
\begin{align}
\Omega_{a,\mathrm{CS}} h^2 \approx 1.6 \times 10^9 f_a^2 m_a^2 \xi \left( -2.25 + \log \frac{f_a}{m_a \xi^{1/2}} \right) 
\times \epsilon^{-1} \left( m_a^2 M_{\mathrm{pl}}^2 g_*^{1/3} \right)^{-3/4}\;,
\end{align}
where, $\xi$ and $\epsilon$ are dimensionless parameters determined by the simulation. 
Based on~\cite{Jia:2024ejr}, their values are approximately $\xi \sim 0.37$ and $\epsilon \sim 0.85$.

In addition, analogously to the effects of the QCD phase transition, the Witten effect contributes to the axion mass and induces the formation of axionic domain walls~\cite{Kawasaki:2015lpf}. 
Axion strings, arising from the breaking of the U(1) PQ symmetry, are attached to axionic domain walls. This configuration is known as a string-wall system, and its decay also contributes to the axion number density~\cite{Lyth:1991bb,Kawasaki:2014sqa}.
The contribution to the DM relic abundance from DFSZ axion~\cite{Shifman:1979if,Zhitnitsky:1980tq} emitted by domain walls, when the domain wall number $N_\text{DW} > 1$, can be described as follows~\cite{Li:2023gil}
\begin{align}
\Omega_{a,\mathrm{DW}} h^2 =& 1.02 \times 10^{-19} 
\left( \frac{8}{3} \right)^{\frac{3}{2p}}
\left( \frac{2p-1}{3-2p} \right) C_d^{\frac{3}{2p}-1}
\left( \frac{m}{\mathrm{GeV}} \right)^{\frac{3}{p} - \frac{3}{2}} \nonumber \\
&\times \left( \frac{f_a}{\mathrm{GeV}} \right)^{4 - \frac{3}{p}}
\Delta V^{1 - \frac{3}{2p}}
\left(\frac{\mathrm{csc}(\pi / \mathrm{N_{Dw}})}{N_{\mathrm{DW}}^2}\right)^{\frac{3}{p}-2}\;,
\end{align}
In calculating the axion relic density, we adopted a method similar to that in ref.~\cite{Li:2023gil}, with parameters $C_d \sim 100$, $p \sim 5/4$, and $N_{\text{DW}} = 3$.

\begin{figure}
    \centering
    \includegraphics[width=0.45\linewidth]{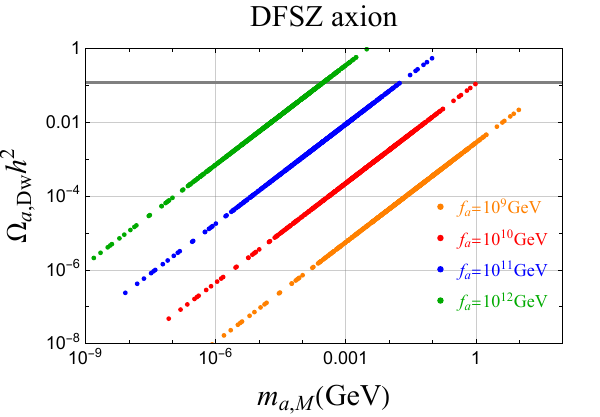}
    \includegraphics[width=0.45\linewidth]{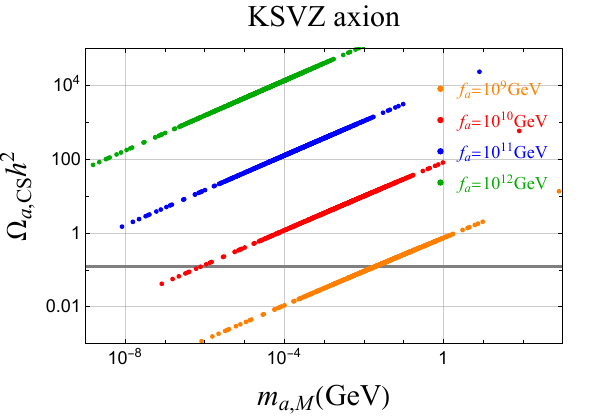}
    \caption{The left (right) panel illustrates the relationship between the abundance of DFSZ axions (KSVZ axions) and the axion mass. The orange, red, blue, and green points represent cases where the axion decay constant $f_a=10^{9}$ GeV, $f_a=10^{10}$ GeV, $f_a=10^{11}$ GeV and  $f_a=10^{12}$ GeV, respectively.}
    \label{fig:axionrd}
\end{figure}
In Fig.~\ref{fig:axionrd}, the dependence of the abundances of the DFSZ and KSVZ axion on the mass $m_{a,M}$ and the decay constant $f_a$ of the axion is presented. It demonstrates a positive correlation between the axion abundance $\Omega_{a}h^2$ and axion mass $m_{a,M}$ and the decay constant $f_a$. The results are shown on the right side of Fig.~\ref{fig:axionrd}, where it is evident that the axion relic density $\Omega_{a,DW}h^2$ decreases as the axion mass $m_{a,M}$ increases. If only the DFSZ axion abundance $\Omega_{a,DW}h^2$ from domain wall decay is considered, the DFSZ axion would need to acquire a larger mass to account for the observed DM relic density.
Furthermore, the contribution of domain wall decay to the DFSZ axion abundance is much smaller than that of cosmic string decay to the KSVZ axion abundance. Current experimental observations of the DM relic density exclude specific parameter ranges of KSVZ axion with larger $f_a$ and heavier $m_{a,M}$.

\section{The gravitational wave and the dark radiation}\label{gw}
The existence of dark phase transitions and domain wall collapse processes suggests that gravitational waves emitted by these events could provide valuable insights into probing new physics models. Our primary focus is on the stochastic gravitational wave background, particularly when its peak frequency lies within the nanohertz range detectable by PTA experiments. Accordingly, we investigate domain wall gravitational waves in the sub-EeV scenario and phase transition gravitational waves in the sub-GeV scenario. Additionally, we examine the constraints imposed by CMB observations on relativistic particles in the dark sector, commonly referred to as dark radiation.

\subsection{Gravitational waves from the dark phase transition}
Evaluating the gravitational wave spectrum produced during a first-order phase transition requires determining the phase transition strength parameter $\alpha$ and the duration parameter $\beta$, defined as follows:
\begin{eqnarray}
\alpha = \frac{\rho_{\text{vac}}}{\rho_{\text{DR}}}, \; \; \; \; 
\beta =  H T \frac{dS_3/T}{dT}, \label{alphabetaForGW}
\end{eqnarray}
where $\rho_{\text{DR}}=\pi^2 g_{\rm DR} T^4/30$ is the plasma energy density, and~\cite{Enqvist:1991xw} the vacuum energy released during the phase transition is defined as below
\begin{eqnarray}
\rho_{\text{vac}} = V_{\text{eff}}(\phi_{\text{phase }1})-V_{\text{eff}}(\phi_{\text{phase }2}) - \frac{T}{4} \frac{\partial}{\partial T} [V_{\text{eff}}(\phi_{\text{phase }1})-V_{\text{eff}}(\phi_{\text{phase }2})] \;.
\end{eqnarray}
Moreover, in subsequent discussions of this work, we perform the calculations at $T_*=T_p$ while adopting the notation $T_*$ for consistency in description.

In this paper, we follow the method outlined in~\cite{Nakai:2020oit} to estimate the gravitational waves generated during the dark first-order phase transition, taking into account contributions from bubble collisions~\cite{Huber:2008hg}, sound waves~\cite{Hindmarsh:2015qta,Hindmarsh:2020hop}, and turbulence~\cite{Caprini:2009yp,Binetruy:2012ze}.
\begin{eqnarray}
\Omega_{\text{GW}}h^2=(\Omega_{\text{bubble}}+\Omega_{\text{sw}}+\Omega_{\text{turb}})h^2.
\end{eqnarray}
where
\begin{align}
 \Omega_{\rm i} h^2 &= \sum_i \Omega_{\rm rad,0} h^2 \left( \frac{g(T_*)}{g_{0}} \right)
 \left( \frac{g_{s0}}{g_{s}(T_*)} \right)^{4/3}
 \left( \frac{H_*}{\beta} \right)^2
 \left( \frac{\kappa_i \alpha'}{1+\alpha'} \right)^2 
  \tilde{\Omega}_{i}. 
  \label{Omega_gw}
\end{align}
Here $\Omega_{\rm rad,0} h^2= 4.16 \times 10^{-5}$~\cite{Fixsen:2009ug} 
with $h$ is the reduced Hubble parameter, and $\alpha^\prime$ can be defined as
\begin{align} 
\label{eq:alpha}
 \alpha' \equiv \frac{\rho_{\rm vac}}{\rho_{\rm rad,tot} (T_*) } =\frac{\rho_{\rm vac}}{\rho_{\rm rad} (T_*) + \rho_{\rm DR} (T_{*})} \simeq \frac{\rho_{\rm vac}}{3 H_*^2 M_p^2}, 
\end{align}
where $\rho_{\rm rad,tot}(T_*)$ is the total radiation energy density just before the phase transition. 
\begin{align}
&\tilde{\Omega}_{\rm bubble}(f) \simeq
    1.0 \,  \left(\frac{0.11 v_w^3}{0.42 + v_w^2}\right)
    \left(\frac{3.8  \left(f/f_{\rm bubble}\right) ^{2.8}}{1+2.8  \left(f/f_{\rm bubble} \right)^{3.8}}\right), \\
&\tilde{\Omega}_{\rm sw}(f) \simeq
    0.16 \, v_w\left(\frac{\beta}{H_*}\right)
    \Upsilon(\bar{U}_f,R_*)
    \left(\frac{f}{f_{\rm sw}}\right)^3\left(\frac{7}{4+3\left(f/f_{\rm sw}\right)^2}\right)^{7/2}, \\
&\tilde{\Omega}_{\rm turb}(f) \simeq
    20 \, v_w\left(\frac{\beta}{H_*}\right)
   \left( \frac{\kappa_{\rm turb} \alpha^\prime}{1+\alpha^\prime} \right)^{-1/2}
    \frac{(f/f_{\rm turb})^3}{(1+(f/f_{\rm turb}))^{11/3}(1+8\pi f/h_*)}, 
\end{align}
where, $f$ is the frequency today, $\Upsilon = (1-1/{\sqrt{1+2 t_{\rm sw} H_*}})$ represents the suppression factor, $\tau_{sw}$ represents the duration of sound wave during the phase transition, as discussed in ref.~\cite{Ellis:2020awk}   $\tau_{sw}=\rm{min}\left[{1}/{H_*}, R_*/{\bar{U}_f}\right]$, \textbf{$H_*R_*=v_w(8\pi)^{1/3}/(\beta/H_*)$}, and the root-mean-square fluid velocity $\bar{U}_f$ can be approximated as \cite{Hindmarsh:2017gnf, Caprini:2019egz, Ellis:2019oqb} $\bar{U}_f^2\approx3\kappa_\nu\alpha^\prime/{4(1+\alpha^\prime)}\;$
and $h_*$ is the Hubble parameter at $T_*$: 
\begin{align}
&h_*\simeq
   1.1\times 10^{-8}\,{\rm Hz}\times\left(\frac{T_*}{0.1\,{\rm GeV}}\right)\left(\frac{g(T_*)}{10.75}\right)^{1/2} \left(\frac{g_{s}(T_*)}{10.75}\right)^{-1/3}\ .
\end{align}
The peak frequencies of three sources $f_{\rm bubble}$, $f_{\rm sw}$, and $f_{\rm turb}$ can be written as:
\begin{align}
&f_{\rm bubble}\simeq
   1.1 \times 10^{-8}\,{\rm Hz}
   \left( \frac{0.62}{1.8 - 0.1 v_w + v_w^2} \right)
   \left(\frac{\beta}{H_*}\right)\left(\frac{T_*}{0.1\,{\rm GeV}}\right)\left(\frac{g(T_*)}{10.75}\right)^{1/2} \left(\frac{g_{s}(T_*)}{10.75}\right)^{-1/3}\  \,,\\
&f_{\rm sw}\simeq
   1.3\times 10^{-8}\,{\rm Hz}\times \frac{1}{v_w} 
   \left(\frac{\beta}{H_*}\right)\left(\frac{T_*}{0.1\,{\rm GeV}}\right)\left(\frac{g(T_*)}{10.75}\right)^{1/2} \left(\frac{g_{s}(T_*)}{10.75}\right)^{-1/3}\  \,,\\
&f_{\rm turb}\simeq
   1.9\times 10^{-8}\,{\rm Hz}\times \frac{1}{v_w}
  \left(\frac{\beta}{H_*}\right)\left(\frac{T_*}{0.1\,{\rm GeV}}\right)\left(\frac{g(T_*)}{10.75}\right)^{1/2} \left(\frac{g_{s}(T_*)}{10.75}\right)^{-1/3}\ \,,
\end{align}

\subsection{Gravitational waves from the domain wall}
To prevent the universe from over-closing, a biased potential $\Delta V$ can be introduced to ensure that domain walls collapse after their formation.
Furthermore, the gravitational wave signals generated during the collapse of these domain walls offer a novel avenue to test this process experimentally. This mechanism provides a new theoretical framework and experimental direction for investigating the properties of cosmological defects and their potential observational signatures. 
The energy released during this process is radiated as gravitational waves, and the peak frequency and peak amplitude of these waves at the present time  $t_0$  can be estimated as follows:~\cite{Hiramatsu:2013qaa,Kadota:2015dza}:
\begin{eqnarray}
f^{dw}\left(t_{0}\right)_{\mathrm{peak}}=\frac{a\left(t_{\mathrm{dec}}\right)}{a\left(t_{0}\right)} H\left(t_{\mathrm{dec}}\right)
\simeq 3.99 \times 10^{-9} \mathrm{Hz} \mathcal{A}^{-1 / 2}\left(\frac{1 \mathrm{TeV}^{3}}{\sigma_{\mathrm{wall}}}\right)^{1 / 2}\left(\frac{\Delta V}{1 \mathrm{MeV}^{4}}\right)^{1 / 2}\;,\label{eq:gwfp} \\
\Omega^{dw}_{\mathrm{GW}} h^{2}\left(t_{0}\right)_{\mathrm{peak}} \simeq 5.20 \times 10^{-20} \times \tilde{\epsilon}_{\mathrm{gw}} \mathcal{A}^{4}\left(\frac{10.75}{g_{*}}\right)^{1 / 3}\left(\frac{\sigma_{\mathrm{wall}}}{1 \mathrm{TeV}^{3}}\right)^{4}\left(\frac{1 \mathrm{MeV}^{4}}{\Delta V}\right)^{2}.\label{eq:gwdw}
\end{eqnarray}
where $\sigma_{wall}\sim c_a m_{a,M} f^2_a$, $c_a = \mathcal{O}(1)$ is a numerical constant. 
The bias term $\Delta V$ introduced in Eqs.~(\ref{eq:gwfp}) and (\ref{eq:gwdw}) explicitly breaks the discrete symmetry, which determines the decay time of the domain wall.
\begin{equation}
t_{dec}\approx \mathcal{A}\sigma_{wall}/(\Delta V)\;.
\end{equation}
Furthermore, another constraint is that domain wall collapse must occur before Big Bang Nucleosynthesis (BBN), that is, $t_{\text{dec}} \leq 0.01 \, \text{sec}$ ~\cite{Kawasaki:2004yh, Kawasaki:2004qu}. This condition places a lower limit on the bias term:
\begin{eqnarray}
 \Delta V \gtrsim 6.6 \times 10^{-2}\mathrm{MeV}^{4} \mathcal{A} \left(\frac{\sigma_{wall}}{1 \mathrm{TeV}^{3}}\right)\;.
\end{eqnarray}
Naturally, the bias term has not only a lower bound but also an upper bound. Specifically, the magnitude of the bias term must be much smaller than the potential around the core of the domain walls ($\Delta V \ll V$), ensuring that the discrete symmetry is approximately preserved.
In this paper, based on the case of  $N_{\text{dw}} = 3$ , we selected the area parameter $\mathcal{A} = 1.2$~\cite{Kadota:2015dza} and the efficiency parameter $\tilde{\epsilon}_{\mathrm{gw}} = 0.7$~\cite{Hiramatsu:2013qaa} , which are consistent with the $\mathbb{Z}_3$ discrete symmetry.
According to the literature~\cite{Hiramatsu:2013qaa}, after calculating the peak frequency and peak amplitude of the gravitational waves, the domain wall gravitational wave power spectrum can be approximated as a piecewise function: for $f < f^{dw}_{\text{peak}} $, $\Omega_{GW}^{\text{dw}} h^2 \propto f^3 $, and for $f \geq f^{dw}_{\text{peak}}$, $\Omega_{GW}^{\text{dw}} h^2 \propto f^{-1}$.

\begin{figure}[htbp]
    \centering
    \includegraphics[width=0.45\linewidth]{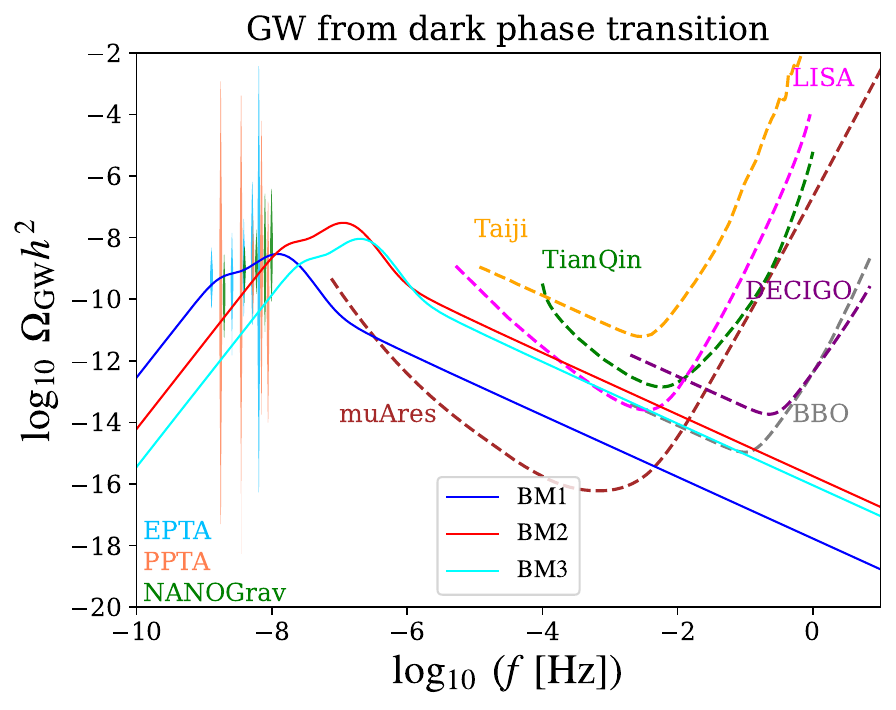}
    \includegraphics[width=0.45\linewidth]{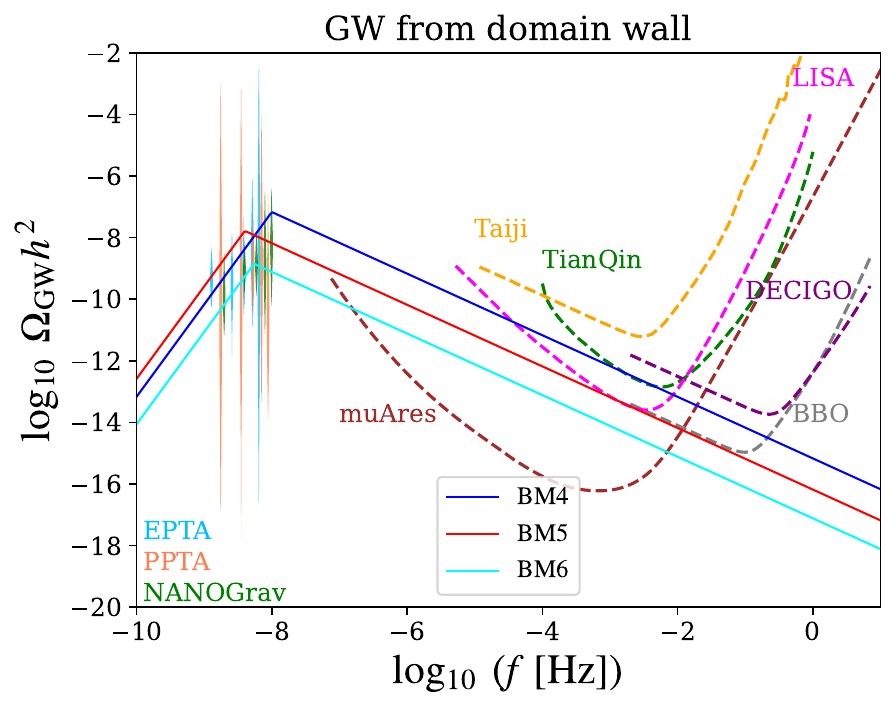}
    \caption{The left panel illustrates the gravitational wave energy spectra generated by dark phase transitions for the first three benchmark points listed in Table\ref{tab:BP}. Meanwhile, the right panel depicts the gravitational wave energy spectra resulting from axion domain wall decay for the last three benchmark points in Table\ref{tab:BP}. The violin plots represent data from three experiments: EPTA~\cite{EPTA:2021crs}, PPTA~\cite{Goncharov:2021oub}, and NANOGrav~\cite{NANOGrav:2020bcs}. Moreover the brown, gray, orange, green, purple and gray dashed lines show the projected sensitivity of the muAres~\cite{Sesana:2019vho}, LISA~\cite{LISA:2017pwj,Baker:2019nia}, Taiji~\cite{Hu:2017mde,Ruan:2018tsw}, TianQin~\cite{TianQin:2015yph}, DECIGO~\cite{Seto:2001qf,Kudoh:2005as,Kawamura:2020pcg} and BBO~\cite{Crowder:2005nr} collaboration, respectively.}
    \label{fig:gw}
\end{figure}
We present the predictions for phase-transition gravitational waves and domain wall collapse gravitational waves at several benchmark points from Table~\ref{tab:BP} in Fig.~\ref{fig:gw}. 
As shown in this figure, the peak frequencies of both types of gravitational waves lie within the nanohertz range, making them detectable by current PTA experiments and future space-based gravitational wave detectors.
\begin{table}[!htbp]
\centering
\begin{tabular}{ccccccc}
\hline\hline
  &  $g_{d} $  &  $\lambda_{\phi}$& $\lambda_{\varphi \phi}$  &  $T_p$~(GeV) & $\beta/H_p$ & $\alpha[T_p]$\\
\hline
$BM_1$  & 0.771 &   0.023  &  $3.91\times10^{-26}$ & 0.001    &  76.47  & 35.57 \\ [+1mm]
$BM_2$  &  0.972 &    0.054  & $1.05\times10^{-22}$ &  0.038 &   23.36 & 114.67 \\ [+1mm]
$BM_3$  &  0.927 &    0.046  & $4.49\times10^{-23}$ &  0.038 &   41.24 & 25.84 \\ [+1mm]
$BM_4$  &  0.903 &    0.064  & $5.407\times 10^{-4}$ & $4.798\times10^{8}$ & 563.75 &  $0.012$   \\ [+1mm]
$BM_5$  &  0.916 &    0.097  & $2.512\times10^{-6}$  & $3.847\times10^{7}$ & 2030.55&   0.0039  \\ [+1mm]
$BM_6$  &  0.833 &    0.049  & $1.875\times10^{-5}$& $9.455\times10^{7}$ & 617.42&  0.013  \\ [+1mm]
\hline \hline
\end{tabular}
\def\baselinestretch{1.1}
\caption{Benchmarks in the Fig.~\ref{fig:gw}. }
\def\baselinestretch{1.0}
\label{tab:BP}
\end{table}
Table~\ref{tab:BP} presents six viable parameter points for the first-order phase transition. The first three correspond to phase transition at the sub-GeV scale, while the last three are associated with sub-EeV scale phase transition.

\subsection{Dark radiation}
This study assumes that the dark sector does not directly couple to the visible sector but is connected through the axion as a mediator particle. In the sub-GeV scenario, the coupling between the dark sector and the axion is extremely weak, implying that the interactions between the dark and visible sectors are negligible. In the sub-EeV scenario, the dark and visible sectors decouple from thermal equilibrium after the axion decouples from the SM. Since the coupling strength between the axion and the SM is significantly suppressed by $f_a$, it can be inferred that the decoupling between the dark sector and the visible sector occurs before the electroweak phase transition. Furthermore, after the first-order phase transition, it is reasonable to assume that the entropy of the dark and visible sectors are separately conserved, which allows us to write the following relation:
\begin{equation}
\frac{g^d_{s}(T_d)T_d^{3}}{g^d_{s}(T_{D,d})T_{D,d}^{3}} = \frac{g^{sm}_{s}(T_{sm})T_{sm}^{3}}{g^{sm}_{s}(T_{D,sm})T_{D,sm}^{3}}.
\end{equation}
where $g^{sm}(T)(g^{sm}_{s}(T))$ and $g^d(T)(g^d_{s}(T))$ are the relativistic degrees of freedom for the energy (entropy) densities in the visible and dark sectors.
When we assume that the first-order phase transition does not occur under supercooling conditions, the extra effective number of relativistic particles can be reasonably expressed as follows \cite{Nakai:2020oit}:
\begin{equation}
\Delta N_{eff} \simeq 0.49 \times \left(\frac{R}{0.13}\right)^{4/3} \left( \frac{g^d_{0}}{g^{sm}_{0}} \right) \left( \frac{g^{sm}_{s,0}}{g^d_{s,0}} \right)^{4/3}\;,
\end{equation}
where $R=s_d(T)/s_{sm}(T)|_{T>T_{pt}}$, the subscript 0 represents the value at the recombination epoch. 

After the breaking of dark symmetry, the dark photon remains massless, whereas the axion mass depending on the specific scenario.
For the sub-EeV dark phase transition, the Witten effect gives the axion a substantial mass, rendering it incompatible with the criteria for dark radiation and leading to $\Delta N_{\text{eff}} = 0.46$.
However, in the case of a sub-GeV phase transition, the axion acquires a much smaller mass from both non-perturbative QCD effect and the Witten effect, which remains far below the temperature $m_a<T_*$. In this scenario, both the dark photon and the axion can be considered as dark radiation, yielding $\Delta N_{\text{eff}}=0.40$. 
Current observations from Planck and BBO provide the constraint~\cite{Planck:2018vyg,riess2018new} $ N_{\text{eff}} = 3.27\pm0.15$. In the SM, $N_{eff}=3.046$.
Moreover, refs.~\cite{Bernal:2016gxb,Planck:2018vyg,Blinov:2020hmc} suggest that $\Delta N_{\text{eff}} \sim 0.4-0.5$ could effectively alleviate tension in the Hubble constant.
\section{Conclusion and Discussions}\label{summary}
In this study, we explore the generation of hidden monopoles through spontaneous breaking of dark $SU(2)_d$ symmetry, focusing on their effects on axion mass and the viability of monopoles and axions as candidates for DM. We also examine the potential of current gravitational wave experiments to detect stochastic gravitational wave backgrounds arising from dark $SU(2)_d$ phase transitions and axionic domain wall collapse.

Our study indicates that in the sub-EeV scenario, the mass of hidden monopoles is around $10^{10}$ GeV, making them viable DM candidates.
Furthermore, the axion mass induced by the Witten effect ranges from $10^{-5}$ GeV to $10^{-1}$ GeV, with its precise value determined by the axion decay constant $f_a$.
The Witten effect contributes significantly more to the axion mass than the non-perturbative QCD effects, making the axion a compelling cold DM candidate. In this scenario, axions radiated by domain walls, their relic density is below the current observation across most of the viable phase-transition parameter space. When the axion decay constant $f_a$ is sufficiently large, such as $10^{11}$ or $10^{12}$ GeV, the DFSZ axion can satisfy the correct relic density. On the other hand, for axions radiated by cosmic strings, their relic density in most of the viable phase-transition parameter space is excluded by the observed relic density. Only when the axion decay constant $f_a$ is sufficiently small, such as $10^9$ or $10^{10}$ GeV, can the KSVZ axion align with current experimental results.
In the sub-GeV scenario, the monopole-induced contribution to the axion mass becomes negligible compared to the contribution from non-perturbative QCD effects. In this case, the axion mass is primarily determined by the QCD instanton effect.

In addition, it is worth noting that in the sub-GeV scenario, the stochastic gravitational wave background from the cosmological first-order phase transition and in the sub-EeV scenario, the stochastic gravitational wave background from axionic domain wall collapse both have peak frequencies in the nano-Hertz range, offering a potential explanation for the low-frequency gravitational wave signals observed by the EPTA, PPTA, and NANOGrav collaborations. This also opens up a new avenue for exploring new physics models that include this dark sector.

Additionally, if the axionic domain wall interacts with lepton or baryon currents, their formation and collapse could facilitate the generation of lepton or baryon number asymmetry, offering a potential explanation for the observed baryon asymmetry in the universe. This mechanism referred to as “DW-genesis” in refs~\cite{Mariotti:2024eoh,Vanvlasselaer:2024vmi}, represents a novel production mechanism. Meanwhile, the contribution of the Witten effect to the axion mass may impact the axiogenesis~\cite{Co:2019wyp,Co:2022aav} process, where the cosmological net baryon number is generated via the rotation of the QCD axion. We leave this topic to future work.

\appendix
\section{Thermal correction} \label{thermal-correction}

The Coleman-Weinberg contribution can be written as \cite{Coleman:1973jx}
\begin{align}
	V_{\rm CW}(\phi)= \sum_{i} \frac{g_{i}(-1)^{b}}{64\pi^2}  m_{i}^{4}(\phi)\left(\mathrm{Log}\left[ \frac{m_{i}^{2}(\phi)}{\Lambda_{UV}^2} \right] - C_i\right)\,,
	\label{eq:oneloop}
\end{align}
where $b=0(1)$ for bosons (fermions), $\Lambda_{UV}$ is the $\overline{MS}$ renormalization scale, $g_i = \{ 1,1,6 \}$ for the $\{ \phi, G_{W^\prime}, W^\prime \}$ in this model,  $C_i = 5/6$ for gauge boson, and $C_i = 3/2$ for others. 
\begin{align}
V_{\rm c.t}= \delta\mu \phi^2 + \delta \lambda \phi^4\;,
\end{align}
where the corresponding coefficients calculated by
\begin{align}
    \frac{\partial V_{\rm c.t}}{\partial \phi} +\frac{\partial V_{\rm CW}}{\partial \phi} |_{\phi \to v_\phi}=0\;,
\frac{\partial^{2} V_{\rm c.t}}{\partial \phi ^2} +\frac{\partial^{2} V_{\rm CW}}{\partial \phi ^2 } |_{\phi \to v_\phi}= 0\;.
\end{align}

The one-loop finite temperature corrections can be written as~\cite{Dolan:1973qd} 
\begin{align}
\label{potVth}
 V_{1}^{T}(\phi, T) = \frac{T^4}{2\pi^2}\, \sum_i n_i J_{B}\left( \frac{ m_i^2(\phi)}{T^2}\right)\;,
\end{align}
where the functions $J_{B}$ are
\begin{align}
\label{eq:jfunc}
J_{B}(y) =  \int_0^\infty\, dx\, x^2\, \ln\left[1 - {\rm exp}\left(-\sqrt{x^2+y}\right)\right]\; ,
\end{align}
with $y\equiv m_{i}^2(\phi)/T^2$\;.

The thermal integrals $J_{B}$ defined in Eq.~(\ref{eq:jfunc}) can be expressed as an infinite series involving modified Bessel functions of the second kind $K_{n} (x)$ with $n=2$~\cite{Anderson:1991zb},
\begin{align}
\label{Bessel_JFJB}
J_{B}(y) = \lim_{N \to +\infty} - \sum_{l=1}^{N}   \frac{y}{l^2}K_{2} (\sqrt{y} l)\;.
\end{align}

The daisy term  $V_{1}^{\rm daisy}(\phi,T)$ is given by\cite{Carrington:1991hz, Arnold:1992rz}
\begin{eqnarray}
V_{1}^{\rm daisy}(\phi,T)=-\frac{T}{12 \pi} \sum_{i={\rm bosons}} n_i \left[ ( m_i^2(\phi)+c_i(T))^{\frac{3}{2}} - ( m_i^2(\phi))^{\frac{3}{2}} \right]\;, \label{DaisyTerms}
\end{eqnarray}
where the finite temperature corrections are given by 
\begin{align}
    c_\phi(T)=\frac{1}{12}T^2 (6g_d^2+5\lambda_\phi) \;\;\;,\;\;\;
    c_{W^\prime}(T)=\frac{11}{6}g_d^2 T^2\;.
\end{align}
\acknowledgments

This work is supported by the National Key Research and Development Program of China under Grant No. 2021YFC2203004.
R.Z. is supported by the National Natural Science Foundation of China under Grant No. 12305109, by Science and Technology Research Project of Chongqing Municipal Education Commission under Grant No.KJQN202300614, and by the National Natural Science Foundation of China under Grant  No. 12147102 
L.B. is supported by the National Natural Science Foundation of China under Grants Nos. 12322505, 12347101, 
L.B. also acknowledges Chongqing Natural Science Foundation under Grant No. CSTB2024NSCQ-JQX0022 and Chongqing Talents: Exceptional Young Talents Project No. cstc2024ycjhbgzxm0020.



\bibliographystyle{JHEP}
\bibliography{biblio.bib}

\end{document}